\documentclass[useAMS,usenatbib,usegraphicx]{mn2e}

\usepackage{multirow}
\usepackage{url}

\newcommand{\grb}{\mbox{GRB\,111215A~}}
\newcommand{\grbnos}{\mbox{GRB\,111215A}}

\title[Afterglow and Host Galaxy of \grbnos]{Detailed Afterglow Modeling and Host Galaxy Properties of the Dark \grbnos}

\author[van der Horst et al.]{A.~J.~van~der~Horst,$^1$\thanks{e-mail: A.J.vanderHorst@uva.nl}, A.J. Levan$^{2}$, G.G. Pooley$^{3}$, K. Wiersema$^{4}$, T. Kr\"{u}hler$^{5,6}$, \newauthor D.A. Perley$^{7,8}$, R.L.C. Starling$^{4}$, P.A. Curran$^{9}$, N.R. Tanvir$^{4}$, R.A.M.J. Wijers$^{1}$, \newauthor R.G. Strom$^{10}$, C. Kouveliotou$^{11}$, O.E. Hartoog$^{1}$, D. Xu$^{12,5}$, J.P.U. Fynbo$^{5}$, \newauthor P. Jakobsson$^{13}$\\
$^1$Anton Pannekoek Institute, University of Amsterdam, Science Park 904, 1098 XH Amsterdam, The Netherlands\\
$^{2}$Department of Physics, University of Warwick, Coventry CV4 7AL, UK\\
$^{3}$Mullard Radio Astronomy Observatory, Cavendish Laboratory, The University of Cambridge, J. J. Thomson Avenue, Cambridge\\ CB3 0HE, UK\\
$^{4}$Department of Physics and Astronomy, University of Leicester, University Road, Leicester LE1 7RH, UK\\
$^{5}$Dark Cosmology Centre, Niels Bohr Institute, University of Copenhagen, Juliane Maries Vej 30, 2100 Copenhagen, Denmark\\
$^{6}$European Southern Observatory, Alonso de C—rdova 3107, Vitacura Casilla 19001, Santiago 19, Chile\\
$^{7}$Department of Astronomy, California Institute of Technology, MC 249-17, 1200 East California Blvd, Pasadena CA 91125, USA\\
$^{8}$Hubble Fellow\\
$^{9}$International Centre for Radio Astronomy Research $-$ Curtin University, GPO Box U1987, Perth, WA 6845, Australia\\
$^{10}$ASTRON, the Netherlands Institute for Radio Astronomy, Postbus 2, 7990 AA Dwingeloo, The Netherlands\\
$^{11}$Space Science Office, ZP12, NASA/Marshall Space Flight Center, Huntsville, AL 35812, USA\\
$^{12}$Department of Particle Physics and Astrophysics, Faculty of Physics, Weizmann Institute of Science, 76100 Rehovot, Israel\\
$^{13}$Centre for Astrophysics and Cosmology, Science Institute, University of Iceland, Dunhagi 5, 107 Reykjav\'{\i}k, Iceland}

\begin{document}

\maketitle

\label{firstpage}

\begin{abstract} 
Gamma-ray burst (GRB) 111215A was bright at X-ray and radio frequencies, but not detected in the optical or near-infrared (nIR) down to deep limits. 
We have observed the GRB afterglow with the Westerbork Synthesis Radio Telescope and Arcminute Microkelvin Imager at radio frequencies, with the William Herschel Telescope and Nordic Optical Telescope in the nIR/optical, and with the {\it Chandra} X-ray Observatory. 
We have combined our data with the {\it Swift} X-Ray Telescope monitoring, and radio and millimeter observations from the literature to perform broadband modeling, and determined the macro- and microphysical parameters of the GRB blast wave. 
By combining the broadband modeling results with our nIR upper limits we have put constraints on the extinction in the host galaxy. 
This is consistent with the optical extinction we have derived from the excess X-ray absorption, and higher than in other dark bursts for which similar modeling work has been performed. 
We also present deep imaging of the host galaxy with the Keck~I telescope, {\it Spitzer Space Telescope}, and {\it Hubble Space Telescope (HST)}, which resulted in a well-constrained photometric redshift, giving credence to the tentative spectroscopic redshift we obtained with the Keck~II telescope, and estimates for the stellar mass and star formation rate of the host. 
Finally, our high resolution {\it HST} images of the host galaxy show that the GRB afterglow position is offset from the brightest regions of the host galaxy, in contrast to studies of optically bright GRBs.
\end{abstract}

\begin{keywords}
gamma-ray bursts: individual: \grb
\end{keywords}

\section{Introduction}\label{sec:intro}

Gamma-ray bursts (GRBs) are observed across the electromagnetic spectrum, and modeling their broadband emission has led to many insights into the physics behind these phenomena. 
The light curves and spectra at various observing frequencies have also provided clues to their immediate environment. 
An example of the latter is the photometry and spectroscopy of GRB afterglows and their host galaxies at optical frequencies, providing not only their redshifts, but also properties of the jets which are producing the emission, and the hosts' gas and dust content. 
There is a sizable fraction of GRB afterglows, however, which are not detected at optical frequencies. 
In fact, 25 to 42 percent of GRBs detected at gamma-ray frequencies by the Burst Alert Telescope (BAT) on-board the {\it Swift} satellite \citep{gehrels2004} are not detected in the optical \citep{fynbo2009,chandra2012}, while for X-ray frequencies this fraction is only $7\%$ \citep{chandra2012}. 
In the early days of GRB afterglow follow-up the fraction of optical non-detections was higher, mainly due to the lack of fast enough response or sensitivity of the available telescopes. 
However, there was a subset of these GRBs with bright X-ray afterglows and deep optical upper limits within hours after the GRB trigger, dubbed dark bursts \citep{groot1998}. 
Possible explanations for their optical darkness are an intrinsic optical faintness or X-ray brightness, a high redshift causing hydrogen absorption in the optical bands, or extinction by gas and dust in their host galaxy \citep[e.g.,][]{djorgovski2001,fynbo2001,rol2005}. 

Different methods have been proposed to classify dark bursts and overcome some of the observational effects in the classification \citep{jakobsson2004,rol2005,vanderhorst2009}. 
While the three proposed methods make different assumptions, they all estimate a minimum expected optical flux based on X-ray spectral or temporal properties. 
Regardless of which classification method one uses, the optical-to-X-ray comparison should be made at sufficiently late times, i.e., at least a few hours after the GRB onset. 
This should be done to exclude intrinsic explanations for the optical darkness, for instance an extra emission component at X-ray frequencies to explain the observed steep decay, shallow decay, and flares at X-ray frequencies \citep[e.g.,][]{nousek2006}. 
These extra emission components are typically prominent for a few hours \citep[e.g.,][]{evans2009,chincarini2010}, after which they not play a significant role, and then either high extinction in the host galaxy or a high redshift due to Lyman-alpha forest absorption are the most likely culprits for the optical darkness. 
If a high redshift can be excluded, for instance by a study of the host galaxy or the near-infrared to optical spectrum, the amount of extinction can be constrained if high quality light curves at X-ray and radio frequencies are available. 
Broadband modeling of the X-ray and radio data, together with upper limits (or faint detections) in the optical bands, can provide lower limits (or estimates) of the optical extinction. 
Since the fraction of GRBs detected at radio frequencies is only one-third \citep{chandra2012}, and detections are regularly at the sensitivity limits of the radio telescopes, the number of GRBs for which this is possible is small. 
The best examples to date, with well sampled light curves at both radio and X-ray frequencies, are GRB\,020819 \citep{jakobsson2005}, GRB\,051022 \citep{rol2007,castrotirado2007}, GRB\,110709B \citep{zauderer2013}, and the focus of this paper: \grb \citep[see also][]{zauderer2013}. 

The BAT onboard the {\it Swift} satellite discovered \grbnos, and a bright X-ray counterpart was found with its X-Ray Telescope \citep[XRT;][]{oates2011a}. 
The burst duration was long, with $T_{90}\,(15-350\,\rm{keV})\,=\,796\pm250$\,s and the source was detected in the BAT for $\sim1500$\,s \citep{barthelmy2011}, which puts it among the longest $\sim1\%$ of GRBs, but still significantly shorter than the ultra-long duration GRBs \citep{levan2014}. 
Possible detections 52 minutes before and 40 minutes after the BAT trigger were reported by MAXI/GSC \citep{kawai2011}, but these were at the 2.7 and $2.2\sigma$ level, respectively, and thus not significant. 
A counterpart of \grb has not been found at near-infrared (nIR), optical or ultraviolet (UV) wavelengths, not within the first minutes to hour \citep{xin2011,xu2011,usui2011,pandey2011,gorbovskoy2011,oates2011b,kuroda2011,rumyantsev2011} nor at later times \citep{aceituno2011,davanzo2011,tanvir2011}. 
However, \grb was detected at several radio frequencies with the Very Large Array \citep[VLA;][]{zauderer2011}, the Combined Array for Research in Millimeter Astronomy \citep[CARMA;][]{horesh2011}, and the IRAM Plateau de Bure Interferometer \citep[PdBI;][]{zauderer2013}, enabling broadband modeling of the radio and X-ray light curves \citep[see also][]{zauderer2013}.

We have observed \grb with the Westerbork Synthesis Radio Telescope (WSRT) at 1.4 and 4.8\,GHz, and with the Arcminute Microkelvin Imager (AMI) at 15\,GHz, resulting in detailed light curves at these three radio frequencies. 
We have also obtained a late-time observation with the {\it Chandra} X-ray Observatory to better constrain the X-ray light curve, and nIR observations of the afterglow with the William Herschel Telescope (WHT) and Nordic Optical Telescope (NOT) to limit the optical darkness of \grbnos. 
We have performed IR-to-optical imaging of the galaxy hosting \grb with the Keck~I telescope, {\it Spitzer Space Telescope}, and {\it Hubble Space Telescope (HST)}, which resulted in a well-constrained photometric redshift, estimates for some physical parameters of the host, and an accurate position of the GRB within the galaxy. 
Our spectroscopy of the host galaxy with the Keck~II telescope resulted in a tentative redshift, consistent with the photometric redshift we have determined. 
In this paper, after presenting our observations of the afterglow (Section~\ref{sec:obs}), and the host galaxy observations and their implications (Section~\ref{sec:host}), we combine our afterglow observations with the radio data from \citet{zauderer2013} to perform broadband modeling (Section~\ref{sec:modeling}), and estimate the physical parameters of the \grb jet and ambient medium. 
We then discuss the optical darkness of \grb in the context of the different classification methods, and put constraints on the optical extinction in the host galaxy based on our modeling work and the deepest nIR observations of the afterglow (Section~\ref{sec:darkness}). 
We summarize our findings in Section~\ref{sec:conclusions}.

\section{Afterglow Observations}\label{sec:obs}

\subsection{Radio Observations}

\begin{table*}
\begin{center}
\caption{Observations with WSRT and AMI of \grbnos, with $\Delta$T the logarithmic midpoint of each observation in days after the trigger time. 
In case of non-detections at 1.4\,GHz, formal flux density measurements for a point source are given, together with upper limits at the 3$\sigma$ level within parentheses.}
\label{tab:wsrt}
\renewcommand{\arraystretch}{1.1}
\begin{tabular}{|l|c|c|c|c|c|} 
\hline
Epoch & $\Delta T$ & Integration & Observatory & Frequency & Flux density \\
 & (days) & time (hours) & & (GHz) & ($\mu$Jy) \\
\hline\hline
2011 Dec 21.489 - 21.953 & 6.13 & 11.1 & WSRT & 4.8 & 467$\pm$37 \\
2011 Dec 25.443 - 25.942 & 10.10 & 5.6 & WSRT & 1.4 & 173$\pm$191 ($<$573) \\
2011 Dec 25.443 - 25.942 & 10.10 & 5.6 & WSRT & 4.8 & 469$\pm$39 \\
2011 Dec 30.429 - 30.928 & 15.09 & 5.6 & WSRT & 1.4 & 237$\pm$175 ($<$525) \\
2011 Dec 30.429 - 30.928 & 15.09 & 5.6 & WSRT & 4.8 & 283$\pm$43 \\
2012 Jan 6.417 - 6.809 & 22.03 & 4.4 & WSRT & 1.4 & 135$\pm$149 ($<$447) \\
2012 Jan 6.417 - 6.809 & 22.03 & 4.4 & WSRT & 4.8 & 1079$\pm$51 \\
2012 Jan 15.386 - 15.814 & 31.01 & 4.8 & WSRT & 1.4 & 373$\pm$170 ($<$510) \\
2012 Jan 15.386 - 15.814 & 31.01 & 4.8 & WSRT & 4.8 & 429$\pm$48 \\
2012 Jan 18.600 - 18.683 & 34.08 & 2.0 & AMI & 15 & 632$\pm$100 \\
2012 Jan 18.710 - 18.850 & 34.19 & 3.3 & AMI & 15 & 791$\pm$70 \\
2012 Jan 19.678 - 19.719 & 35.12 & 1.0 & AMI & 15 & 473$\pm$70 \\
2012 Jan 21.728 - 21.770 & 37.18 & 1.0 & AMI & 15 & 955$\pm$80 \\
2012 Jan 22.544 - 22.585 & 37.97 & 1.0 & AMI & 15 & 685$\pm$70 \\
2012 Jan 23.713 - 23.754 & 39.15 & 1.0 & AMI & 15 & 709$\pm$70 \\
2012 Jan 25.522 - 25.564 & 40.97 & 1.0 & AMI & 15 & 865$\pm$70 \\
2012 Jan 26.479 - 26.520 & 41.93 & 1.0 & AMI & 15 & 712$\pm$90 \\
2012 Jan 28.605 - 28.646 & 44.05 & 1.0 & AMI & 15 & 588$\pm$70 \\
2012 Jan 29.503 - 29.836 & 45.08 & 8.0 & AMI & 15 & 576$\pm$45 \\
2012 Jan 30.518 - 30.551 & 45.96 & 0.8 & AMI & 15 &  513$\pm$70 \\
2012 Feb 1.339 - 1.838 & 48.00 & 5.6 & WSRT & 1.4 & 199$\pm$167 ($<$501) \\
2012 Feb 1.339 - 1.838 & 48.00 & 5.6 & WSRT & 4.8 & 461$\pm$42 \\
2012 Feb 1.465 - 1.798 & 48.04 & 8.0 & AMI & 15 & 595$\pm$35 \\
2012 Feb 9.683 - 9.724 & 56.12 & 1.0 & AMI & 15 & 509$\pm$70 \\
2012 Feb 11.710 - 11.752 & 58.15 & 1.0 & AMI & 15 & 511$\pm$70 \\
2012 Feb 25.274 - 25.772 & 71.94 & 12.0 & WSRT & 4.8 & 394$\pm$34 \\
2012 Feb 25.392 - 25.725 & 71.98 & 8.0 & AMI & 15 & 411$\pm$35 \\
2012 Feb 27.397 - 27.439 & 73.85 & 1.0 & AMI & 15 & 341$\pm$140 \\
2012 Mar 1.629 - 1.671 & 77.07 & 1.0 & AMI & 15 & 335$\pm$70 \\
2012 Mar 3.555 - 3.597 & 79.00 & 1.0 & AMI & 15 & 151$\pm$70 \\
2012 May 6.080 - 6.579 & 142.74 & 5.6 & WSRT & 1.4 & 102$\pm$151 ($<$453) \\
2012 May 6.080 - 6.579 & 142.74 & 5.6 & WSRT & 4.8 & 149$\pm$41 \\
2012 Aug 7.823 - 8.322 & 236.49 & 12.0 & WSRT & 1.4 & 39$\pm$34 ($<$102) \\
2012 Aug 9.818 - 10.316 & 238.48 & 12.0 & WSRT & 4.8 & 76$\pm$31 \\
\hline
\end{tabular}
\end{center}
\end{table*}

In our WSRT observations at 1.4 and 4.8\,GHz we used the Multi Frequency Front Ends \citep{tan1991} in combination with the IVC+DZB back end in continuum mode, with a bandwidth of 8x20 MHz at both observing frequencies. 
Complex gain calibration was performed with the calibrator 3C\,286 for all observations. 
The observations were analyzed using the Multichannel Image Reconstruction Image Analysis and Display \citep[MIRIAD;][]{sault1995} package. 

We observed \grb with the Large Array (LA) of AMI \citep{zwart2008} using a bandwidth of 3.75\,GHz around a central frequency of 15.4\,GHz. 
J2321+3204 was observed at regular intervals for phase calibration, and the flux density scale was established by observations of 3C\,48 and 3C\,286. 
The data were analysed using the in-house software package \textsc{reduce}. 
The details and results of our WSRT and AMI observations are given in Table~\ref{tab:wsrt}.

\subsection{Near-Infrared Observations}

We observed the field of \grb using the Long-slit Intermediate Resolution Infrared Spectrograph (LIRIS) mounted in the Cassegrain focus of the 4.2m WHT at La Palma, Spain. 
We acquired imaging in the $J$ and $K_s$~bands, under poor conditions, with average seeing of 2.0\arcsec. 
A total exposure time of 648\,s in $K_s$~band and 630\,s in $J$~band was used, at mid-times after the GRB onset of 4.92 and 5.25\,h, respectively. 
We reduced the data using the \textsc{lirisdr} data reduction package (within IRAF), which has been developed by the LIRIS team and performs several LIRIS-specific corrections in addition to the standard reduction steps. 
We calibrated astrometry and photometry using 2MASS sources in the field. 
No source was detected at or near the afterglow position, and we derived 3$\sigma$ limits of $K_s>19.2$ and $J > 19.5$ (Vega magnitudes). 

We also observed the field of \grb with the Nordic Optical Telescope equipped with StanCam, starting on December 15.847, 2011. 
In total six images with an integration time of 600\,s each were obtained in the $z'$ band. 
The data were reduced using standard techniques in pyraf/IRAF, and our photometric calibration was done with local field stars from DR8 of the SDSS catalog \citep{aihara2011}. 
In the stacked image, no source was detected at the afterglow position down to a 3$\sigma$ limiting AB magnitude of $z'>22.6$ at a mid-time of 6.83\,h after the GRB onset.

\subsection{X-Ray Observations}

We observed \grb with the Advanced CCD Imaging Spectrometer (ACIS-S) onboard the {\it Chandra} X-ray Observatory. 
The observation started on 28 December 2011 at 04.79 UT and lasted for 15\,ks, i.e., the observation midpoint was 12.7\,d after the burst onset. 
\grb was detected and the spectrum adequately modeled by an absorbed power law, with photon index $\Gamma=1.57$, and an excess absorption of $N_{\rm{H,exc}}=3.8\times10^{21}$\,cm$^{-2}$ compared to the Galactic value of $N_{\rm{H,gal}}=6.5\times10^{20}$\,cm$^{-2}$ \citep{willingale2013}. 
We measure an unabsorbed flux of $(1.59\pm0.14)\times10^{-13}$\,erg\,s$^{-1}$\,cm$^{-2}$ (0.3$-$10\,keV), and a 2\,keV flux of $(6.5\pm0.6)\times10^{-6}$\,mJy. 

{\it Swift} X-Ray Telescope (XRT) light curves and spectra were obtained from the UKSSDC online Repository\footnote{\url{www.swift.ac.uk/xrt_products}} \citep{evans2009}. 
We have determined the 2\,keV flux for each observation by assuming that the photon index is constant at $\Gamma=2.00\pm0.10$, based on the time-averaged Photon Counting mode spectrum. 
The resulting X-ray light curve, including the {\it Chandra} data point, is shown in the top left panel of Figure~\ref{fig:lcs}.

\section{Host Galaxy}\label{sec:host}

\subsection{Keck and Spitzer Space Telescope Imaging}\label{sec:keckspitzer}

We imaged the field of \grb at several epochs between 26 December 2011 and 2013 June 20 with the Keck~I telescope equipped with the Low Resolution Imaging Spectrometer \citep[LRIS;][]{oke1995} or the Multi Object Spectrograph for Infra Red Exploration \citep[MOSFIRE;][]{mclean2012}. 
Images were obtained in 8 different filters from the $u$ to the $K_s$ band, and the host galaxy was significantly detected in all of them. 
Our Keck imaging was reduced entirely with a custom made pipeline based on IDL using standard techniques for CCD or nIR data, depending on the observing band.

We used standard aperture photometry by choosing an aperture size and location aided by our {\it HST} imaging with a minimum radius of 1\farcs1, that encompasses the full visible extent of the galaxy in the {\it HST} image (see Section~\ref{sec:hst}). 
Photometric calibration was then performed against field stars from the SDSS \citep{aihara2011} or the 2MASS \citep{skrutskie2006} catalog in the respective filter bands. 
For observations in filters that are not covered by the respective surveys, e.g., $R$ or $Y$-band, we used synthetic photometry of stellar spectra to derive brightnesses of field stars from bracketing pass-bands (i.e., $r$ and $i$ for $R$, or $z$ and $J$ for $Y$). 

Observations of the host of \grb with the Infrared Array Camera \citep[IRAC;][]{fazio2004} onboard {\it Spitzer} \citep{werner2004} in its Warm Mission were taken on 31 January 2013. 
The observing strategy and analysis follow closely what has been reported in \citet{perley2013}, with 16 dithered exposures of $100\,s$ integration time in both channel~1 (central wavelength: 3.6\,$\mu$m) and channel~2 (central wavelength: 4.5\,$\mu$m). 
We downloaded the PBCD files from the Spitzer Legacy Archive\footnote{\url{archive.spitzer.caltech.edu}}, measuring the magnitude of the host using circular aperture photometry and calibrated via the zeropoints given in the IRAC handbook\footnote{\url{irsa.ipac.caltech.edu/data/SPITZER/docs/irac/iracinstrumenthandbook}}. 

Details and results of our Keck and {\it Spitzer} photometry are given in Table~\ref{tab:hostmags}.

\begin{table}
\begin{center}
\caption{Photometry of the host galaxy of \grb with Keck, {\it Spitzer} and {\it HST}. 
All photometry is given in the AB system, and is not corrected for Galactic foreground reddening.}
\label{tab:hostmags}
\renewcommand{\arraystretch}{1.1}
\begin{tabular}{|l|c|c|c|}
\hline
Filter & Instrument & Epoch & Magnitude \\
\hline\hline
$u$ & Keck/LRIS & 2012 Dec 11 & $25.40\pm0.20$ \\
$g$ & Keck/LRIS & 2011 Dec 26 & $24.60\pm0.05$ \\
$g$ & Keck/LRIS & 2012 Jan 26 & $24.62\pm0.07$ \\
F606W & {\it HST}/WFC3 & 2013 May 13 & $24.50\pm0.13$ \\
$R$ & Keck/LRIS & 2011 Dec 26 & $24.25\pm0.06$ \\
$i$ & Keck/LRIS & 2012 Jan 26 & $24.08\pm0.08$ \\
$z$ & Keck/LRIS & 2012 Dec 11 & $23.67\pm0.17$ \\
$Y$ & Keck/MOSFIRE & 2013 Jun 20 & $23.70\pm0.16$ \\
$J$ & Keck/MOSFIRE & 2013 Jun 20 & $22.79\pm0.15$ \\
F160W & {\it HST}/WFC3 & 2013 May 13 & $22.61\pm0.03$ \\
$H$ & Keck/MOSFIRE & 2013 Jun 20 & $22.24\pm0.17$ \\
$K_{\rm{s}}$ & Keck/MOSFIRE & 2013 Jun 20 & $21.95\pm0.14$ \\
3.6$\,\mu$m & $Spitzer$/IRAC & 2013 Jan 31 & $21.70\pm0.06$ \\
4.5$\,\mu$m & $Spitzer$/IRAC & 2013 Jan 31 & $21.51\pm0.20$ \\
\hline
\end{tabular}
\end{center} 
\end{table}

\subsection{Keck Spectroscopy}

\begin{figure}
\begin{center}
\includegraphics[angle=0,width=\columnwidth]{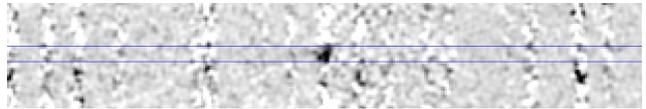}
\caption{Sky-subtracted, co-added near-infrared spectra from NIRSPEC on Keck II showing the possible H-alpha line at 1.977\,$\mu$m ($z$=2.012) at the center of the image. 
This line falls on top of a sky emission feature, but is located directly at the position of the host galaxy, labeled in blue, and corresponds to a redshift consistent with the narrow range derived photometrically (see Section \ref{sec:photoz}).}
\label{fig:spectrum}
\end{center}
\end{figure}

We acquired spectroscopy of the host galaxy of \grb on 24 August 2012 using the Near-Infrared Spectrograph \citep[NIRSPEC;][]{mcclean1998} on Keck~II. 
Our observations consisted of four 900\,s exposures in NIRSPEC-1 configuration ($0.95-1.12\,\mu$m), four 600\,s exposures in the NIRSPEC-3 configuration ($1.14-1.37\,\mu$m), and two 600\,s exposures in the NIRSPEC-6 configuration ($1.75-2.18\,\mu$m).

We carefully inspected the individual sky-subtracted exposures in all three configurations for emission lines. 
No credible candidates were found in the NIRSPEC-1 or NIRSPEC-3 setups. 
In the NIRSPEC-6 observations we identified a probable line at 1.977\,$\mu$m; it is present in both exposures (although only marginally significant in the second exposure) and on top of a bright night-sky line; see Figure~\ref{fig:spectrum} for the sky-subtracted, co-added 2D~spectrum around the putative line. 
Given photometric constraints on the redshift (Section~\ref{sec:photoz}), the only strong line candidate matching this wavelength is H$\alpha$ at $z=2.012$. 
This association is consistent with the nondetection of lines in our other configurations. 
In particular, [OII] at that redshift would lie in the narrow coverage gap between NIRSPEC-1 and NIRSPEC-3 observations (i.e., at 1.123\,$\mu$m), and [OIII] would lie at the blue end of the $H$-band (at 1.315\,$\mu$m), also outside our coverage. 
This redshift is also consistent with the nondetection of emission lines in the optical spectroscopy of \citet{zauderer2013}. 
However, with only a single line detected, and only in two exposures, this redshift must be regarded as tentative.

\subsection{Host Galaxy Photometric Redshift and Physical Parameters}\label{sec:photoz}

\begin{figure}
\begin{center}
\includegraphics[angle=0,width=\columnwidth]{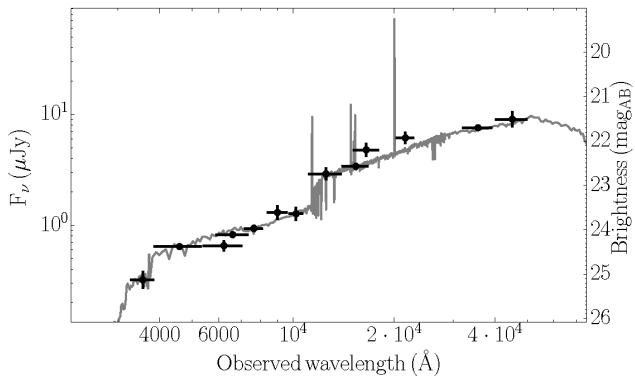}
\caption{Photometry of the galaxy hosting \grb and the best-fit stellar population synthesis model. 
The black data points indicate the photometry from Table~\ref{tab:hostmags}, corrected for a Galactic foreground reddening of $E_{B-V}=0.055\,\rm{mag}$, and the grey curve represents the best-fit model.}
\label{fig:photoz}
\end{center}
\end{figure}

Since the optical afterglow of \grb was not detected, there is no redshift based on afterglow spectroscopy. 
\citet{zauderer2013} put constraints on the allowed redshift range based on the lack of emission lines in the spectrum of the host galaxy: $1.8\leq z\leq2.7$. 
Here we derive a better constrained, photometric redshift, based on our IR-to-optical imaging of the host galaxy. 
We fit the available photometry presented in Table~\ref{tab:hostmags} with a stellar population synthesis model within LePHARE \citep{arnouts1999}. 
We created a model grid of $\sim10^{6}$ galaxy spectra with different star-formation histories, ages, metallicities and dust attenuation properties \citep{calzetti2000}, using templates from \citet{bruzual2003}. 
Emission lines were added to the stellar continuum by using the ongoing star-formation rate of the galaxy template via the descriptions of \citet{kennicutt1998}.

We obtained the best model fit for a redshift of $z_{\rm{phot}}=2.06^{+0.10}_{-0.16}$, with $\chi^2=15$ for 13 filters. 
The redshift solution is driven by the prominent Balmer break between the $Y$ and $J$ filters, as well as the $u-g$ color that is consistent with the onset of the Lyman forest at $z\sim2$ (see Figure~\ref{fig:photoz}). 
This redshift makes emission line spectroscopy very challenging even in the nIR, because prominent emission lines such as H$\alpha$ are likely located in the atmospheric absorption bands.

We used the same SED fit to put constraints on some physical parameters of the host. 
The galaxy hosting \grb has a stellar mass of $\log\rm{(M_{*}/{M_{\sun})}}={10.5}^{+0.1}_{-0.2}$ and a luminosity of $M_B=-22.0\pm0.2$. 
A mild dust-correction of $\rm{E_{B-V}} \sim 0.15\,\rm{mag}$ yields a dust-corrected star-formation rate $\rm{SFR_{SED}} = 34^{+33}_{-13}\,\rm{M_{\sun}\,yr^{-1}}$. 
These properties are similar to galaxies selected through GRBs with a highly extinguished optical afterglow \citep{kruhler2011,perley2013}.

\subsection{Hubble Space Telescope Imaging and the GRB Position in its Host Galaxy}\label{sec:hst}

\begin{figure}
\begin{center}
\includegraphics[angle=90,width=0.95\columnwidth]{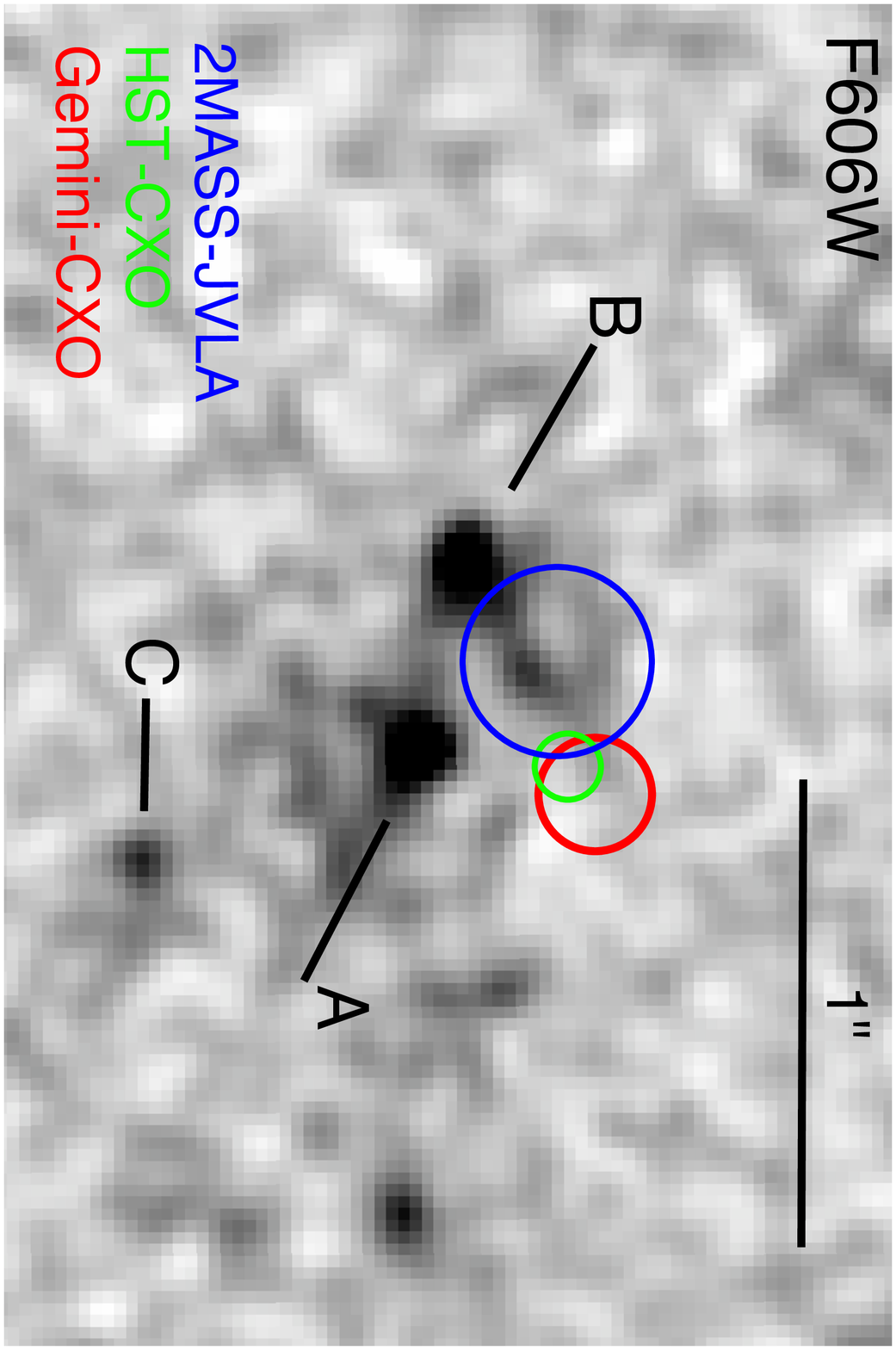}
\includegraphics[angle=90,width=0.95\columnwidth]{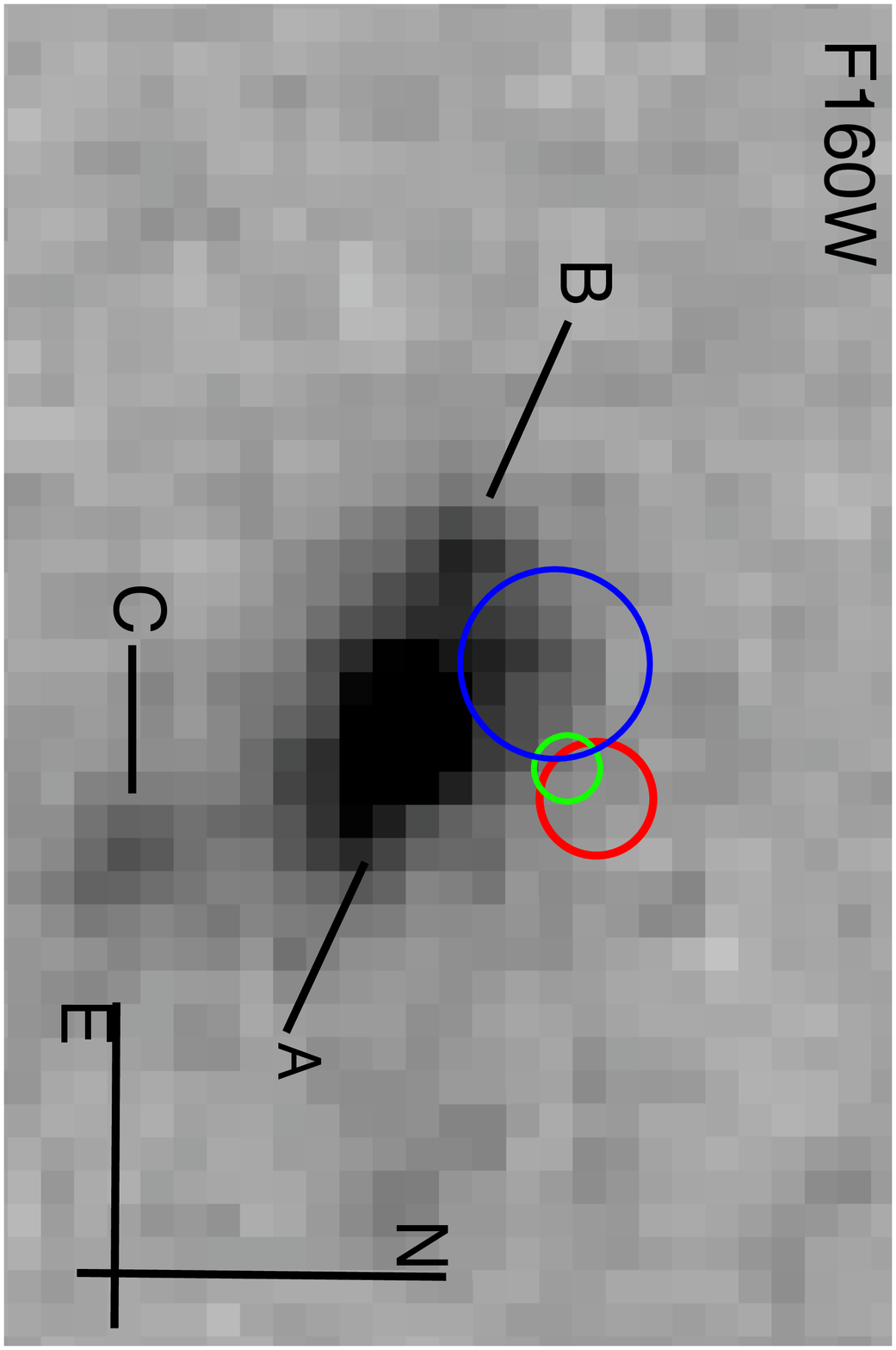}
\caption{{\it HST} images of the \grb host galaxy complex in the optical (F606W; top panel) and nIR (F160W; bottom panel). 
The brightest spots in the complex are indicated by A, B and C. 
The three colored circles show the \grb position and uncertainty obtained through the three different methods discussed in Section~\ref{sec:hst}: using 2MASS and the VLA (left blue circle); {\it HST} and {\it Chandra} (middle green circle); and {\it Chandra} and Gemini (right red circle).}
\label{fig:host}
\end{center}
\end{figure}

The \grb field was observed with the {\it Hubble Space Telescope (HST)} on 13 May 2013. 
We obtained observations in the optical (UVIS filter F606W; central wavelength: 5887\,\AA) with an exposure time of 1110\,s and nIR (IR filter F160W; central wavelength: 15370\,\AA) for 1209\,s with the Wide Field Camera~3. 
Data were reduced using \textsc{astrodrizzle} in the standard fashion, with a final pixel scale of 0.025\arcsec\,pixel$^{-1}$ for F606W and 0.07\arcsec\,pixel$^{-1}$ for F160W. 
The host environment, as seen in the higher resolution F606W image in Figure~\ref{fig:host}, is clearly complex, consisting of multiple knots of bright (likely star forming) emission, as well as fainter, low surface brightness regions. 
This is consistent with an interacting or possibly merging environment, which could have triggered the star formation responsible for the GRB. 
One of these regions (marked A in Figure~\ref{fig:host}) is significantly brighter in the nIR than in the optical, suggesting the presence of either a significantly obscured component or a much older stellar population. 
The global magnitudes of the host complex are F606W(AB)=24.50$\pm$0.13 and F160W(AB)=22.61$\pm$0.03 (see also Table~\ref{tab:hostmags}), with the brightest knot (A) showing F606W(AB)=26.15$\pm$0.09 and F160W(AB)=23.41$\pm$0.03. 
These colours are red, especially for A, and are comparable to those of Extremely Red Objects, which remain relatively rare amongst GRB hosts, although they are more frequently represented in the dark-GRB population \citep[e.g.,][]{levan2006,berger2007,hjorth2012,rossi2012,perley2013}.

We used multiple approaches to ascertain the position of \grb within its host complex. 
Firstly, we directly aligned the {\it Chandra} images with the UVIS and IR fields of view. 
In this case we obtained a total of 3 X-ray sources within the field, including the GRB afterglow. 
The other two sources appeared as resolved galaxies, and we assumed that the X-ray sources within them are nuclear; although redshifts are not available for these galaxies, their magnitudes and angular extent are such that we would not expect to observe contributions from discrete source populations within them, but only from AGN. 
We determined the expected positions of these two sources with the world-coordinate system (WCS) of the {\it HST} observations, and calculated the shift in position to map to the centroids of the sources in the {\it HST} images. 
We then determined the afterglow position by applying the same shift to the X-ray afterglow of \grbnos, which gives a positional offset between the two sources of $\sim0.065$\arcsec. 
This was added in quadrature to the statistical error in the position of the afterglow in the {\it Chandra} observations (FWHM/2.3$\times$S/N$\approx$0.019\arcsec) to provide a final position accuracy of $\sim0.07$\arcsec. 
The positions were determined from the centres of the sources as measured in both the IR and UVIS frames to allow for any shifts due to extinction, and each position is consistent with the other. 
However, the small number of sources used in this approach precludes a detailed fit to the data, and so may suffer from an additional systematic error. 
Hence we also investigated two additional routes to placing the GRB on its host. 
The first was to use 6 2MASS sources within the UVIS field of view to attempt to fix the {\em HST} imaging to a fixed WCS, onto which we can place the well localised VLA position. 
The scatter of this was relatively large ($\sim 0.2$ arcsec), which is likely due to the difficulty in centroiding the saturated bright stars, as well as their proper motion in the time frame between 2MASS and {\em HST} observations. 
Finally, we used a Gemini-N observation of the field of GRB 111215A \citep[also reported in][]{zauderer2013} to provide a wider field of view for the identification of optical counterparts to {\em Chandra} detected X-ray sources. 
This approach identified 8 objects in common to the Gemini and {\em Chandra} frames, providing a relative match between the two of 0.12\arcsec; the match between the Gemini and UVIS observations has a scatter of only 0.03\arcsec and so does not contribute significantly to the overall error budget. 
The error regions determined by these three different approaches all overlap at the 1$\sigma$-level, and are shown graphically in Figure~\ref{fig:host}.

Interestingly, the position of the X-ray afterglow lies offset from any bright regions of star formation based on any of the above routes to astrometry, which would suggest it is not associated with the strongest regions of star formation in its host, in contrast to most optically bright bursts \citep[e.g.,][]{fruchter2006,svensson2010}. 
It is possible that the progenitor formed in a region of less intense star formation, or that its birth region is so obscured that it is not visible even in the nIR (the rest-frame optical given our photometric redshift).

\section{Broadband Modeling}\label{sec:modeling}

\begin{figure*}
\begin{center}
\includegraphics[angle=-90,width=0.95\textwidth]{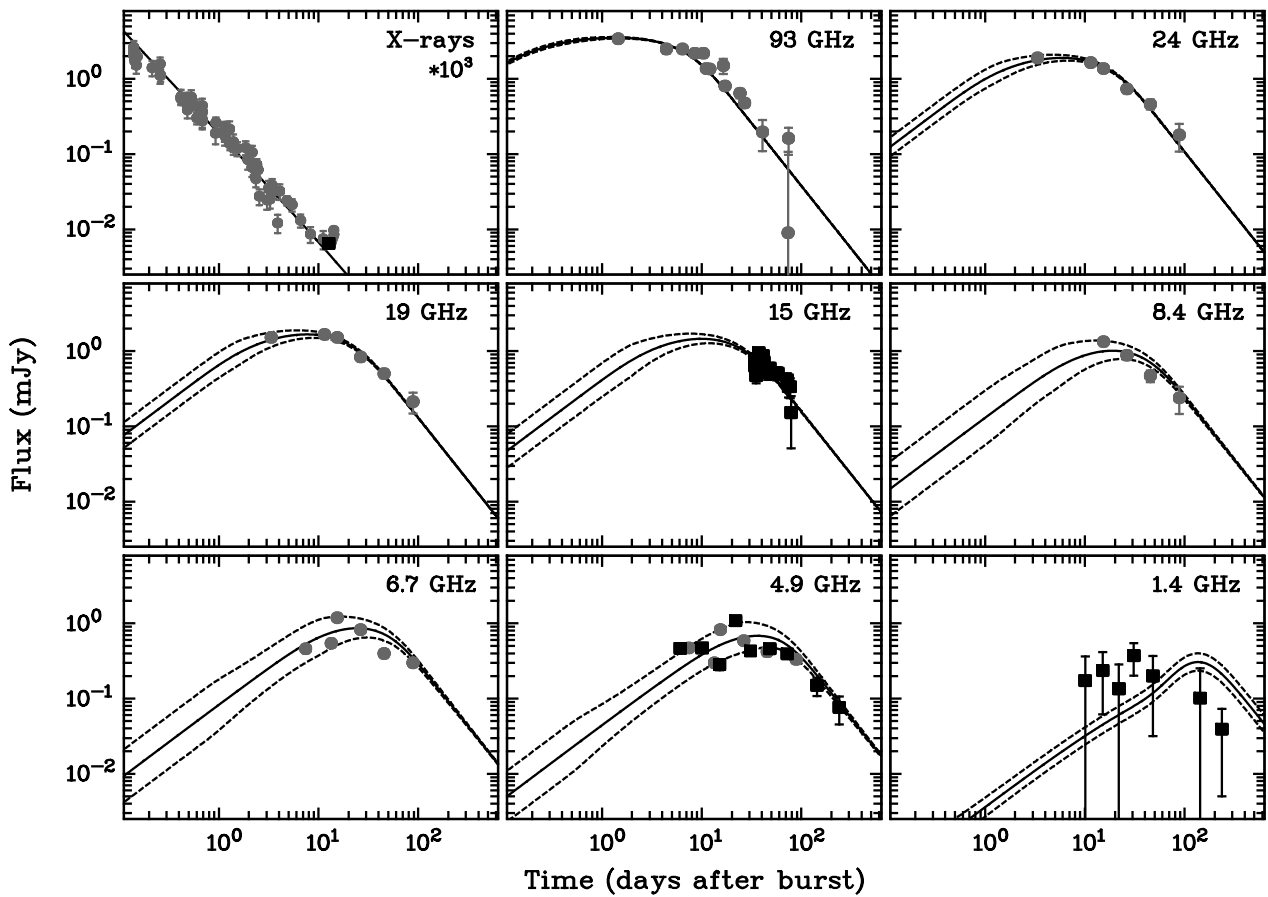}
\caption{Broadband light curves with the best fit model. 
The X-ray light curve shows the flux density evolution at 2\,keV, multiplied by a factor $10^3$, with the grey circles indicating the {\it Swift}/XRT flux densities and the black square the {\it Chandra} data point. 
The radio and millimeter data have been obtained with the WSRT and AMI (black squares; this paper), and with the VLA, CARMA and PdBI \citep[grey circles;][]{zauderer2013}. 
The solid lines correspond to the best fit broadband model, and the dashed lines indicate the predicted scatter in the radio bands due to interstellar scintillation.}
\label{fig:lcs}
\end{center}
\end{figure*}

We have combined the radio and X-ray data presented in Section~\ref{sec:obs} with the radio and millimeter data presented in \citet{zauderer2013}, spanning a $5-93$\,GHz frequency range, to model the \grb afterglow. 
We performed broadband modeling in the standard framework of a relativistic blast wave at the front of a jet, in which relativistic electrons are emitting broadband synchrotron emission. 
The broadband spectrum can be characterized by three break frequencies \citep{sari1998}: the peak frequency $\nu_{\rm{m}}$, the cooling frequency $\nu_{\rm{c}}$, and the synchrotron self-absorption frequency $\nu_{\rm{a}}$. 
The evolution of these characteristic frequencies is governed by the dynamics of the blast wave, and determines the light curves at given observing frequencies. 
From these three frequencies and the peak flux $F_{\rm{\nu,max}}$ one can determine four physical parameters: two parameters describing the macrophysics, namely the isotropic equivalent kinetic energy $E_{\rm{K,iso}}$ of the blast wave and the density $\rho$ of the ambient medium; and two microphysics parameters, i.e., the fractions $\varepsilon_{\rm{e}}$ and $\varepsilon_{\rm{B}}$ of the internal energy density in electrons and the magnetic field, respectively. 
Besides these four parameters there are three more that shape the broadband spectra and light curves: the power-law index $p$ of the energy distribution of the emitting electrons, the power-law index $k$ of the ambient medium density with radius ($\rho=A\cdot R^{-k}$), and the jet opening angle $\theta_{\rm{j}}$. 
In principle there are two more parameters, the observing angle $\theta_{\rm{obs}}$ and the fraction $\xi$ of electrons participating in the relativistic power-law energy distribution. 
These two parameters, however, cannot be constrained by the \grb data set, so we make the typical assumptions $\theta_{\rm{obs}}=0$ and $\xi=1$. 

We only take the emission from the forward shock moving into the ambient medium into account, and we do not model a possible contribution from the reverse shock moving into the jet. 
With the long prompt emission duration of \grb one could consider if the X-ray afterglow is possibly dust scattered prompt emission, as was invoked for the ultra-long GRB\,130925A \citep{evans2014}. 
This not the case for \grb because (i) the prompt emission is not very bright compared to the X-ray afterglow, and (ii) the X-ray spectrum does not show softening over time. 
In the following we show that the X-ray, radio and millimeter light curves can indeed be fit well with emission from the forward shock without any dust scattering or a reverse shock component.

\subsection{X-Ray Light Curve}

Before performing a full broadband fit of all the available data, we first focus on the X-ray spectrum and light curve to constrain some of the parameters. 
The observed X-ray spectral slope $\beta=1.00\pm0.10$ and temporal slope $\alpha=1.32\pm0.02$ should satisfy the so-called closure relations, i.e., the derived values for $p$ from $\beta$ and $\alpha$ should be consistent for the same spectral regime and value of $k$. 
Since the spectral slope is $(p-1)/2$ for frequencies in between $\nu_{\rm{m}}$ and $\nu_{\rm{c}}$, and $p/2$ above $\nu_{\rm{c}}$, the inferred value for $p$ is $3.00\pm0.20$ for $\nu_{\rm{m}}<\nu_{\rm{X}}<\nu_{\rm{c}}$ or $2.00\pm0.20$ for $\nu_{\rm{m}}<\nu_{\rm{c}}<\nu_{\rm{X}}$. 
The temporal slopes depend on the structure of the circumburst medium, for which we investigate two specific cases: a homogeneous medium ($k=0$) and a stellar wind with a constant velocity ($k=2$). 
Based on our observed temporal slope we derive the following $p$ values \citep{sari1998,chevalier1999}: $p=2.76\pm0.03$ for $\nu_{\rm{m}}<\nu_{\rm{X}}<\nu_{\rm{c}}$ and $k=0$; $p=2.09\pm0.03$ for $\nu_{\rm{m}}<\nu_{\rm{X}}<\nu_{\rm{c}}$ and $k=2$; and $p=2.43\pm0.03$ for $\nu_{\rm{m}}<\nu_{\rm{c}}<\nu_{\rm{X}}$ independent of the value of $k$. 
Comparing the $p$ values derived from $\alpha$ and $\beta$ implies that $\nu_{\rm{m}}<\nu_{\rm{X}}<\nu_{\rm{c}}$ for $k=0$ is preferred, but $\nu_{\rm{m}}<\nu_{\rm{c}}<\nu_{\rm{X}}$ (for all $k$) is also consistent within $2\sigma$. 

Another parameter constraint we can obtain from the X-ray light curve is a lower limit on the so-called jet-break time $t_{\rm{j}}$, when the blast wave has decelerated to a Lorentz factor $\Gamma$ for which $\theta_{\rm{j}}=1/\Gamma$. 
After $t_{\rm{j}}$ the X-ray light curve is expected to show a significant steepening to a slope equal to $-p$ \citep{rhoads1999}. 
The X-ray light curve can be fit well with a single power law, resulting in $\chi^2=91.8$ and $\chi^2_{\rm{red}}=1.4$. 
A fit including a jet break as a free parameter puts $t_{\rm{j}}$ above the observed time range, especially because of our well constrained {\it Chandra} data point. 
We can estimate a lower limit on the jet break time by forcing a sharp break at $t_{\rm{j}}$ inside the observed time range, and determine at which value the $\chi^2$ becomes significantly worse while $p$ and the normalization are free parameters. 
In this particular case we require $\Delta\chi^2=14.2$ for a $3\sigma$ limit, and we find $t_{\rm{j}}>8.8$\,d.

\subsection{Broadband Light Curves}

We have modeled the radio and X-ray light curves of \grb using the methods and broadband fitting code from \citet{vanderhorst2007}. 
For the dynamics of the blast wave this method adopts the \citet{blandford1976} solutions at early times, and an evolution following \citet{rhoads1999} after the jet-break time. 
Fits with $k$ as a free parameter did not converge to stable parameter solutions, and thus we have explored $k=0$ and $k=2$ for the ambient medium density structure. 
First we fit the light curves with a spherical blast wave model, and we obtained the best fit for $k=2$, with $\chi^2=1196$ and $\chi^2_{\rm{red}}=6.2$, and $\nu_{\rm{m}}<\nu_{\rm{c}}<\nu_{\rm{X}}$ for the entire observed time range. 
Fits with $k=0$ resulted in $\chi^2=2339$ and $\chi^2_{\rm{red}}=12.1$ in the spherical case; adding a jet break improved the fits slightly, although with $\chi^2=2021$ and $\chi^2_{\rm{red}}=10.5$ still worse than the $k=2$ spherical blast wave fits. 
We note that our analysis of the X-ray light curve and spectrum resulted in $\nu_{\rm{m}}<\nu_{\rm{X}}<\nu_{\rm{c}}$ with $k=0$ being preferred, but that $\nu_{\rm{m}}<\nu_{\rm{c}}<\nu_{\rm{X}}$ is also consistent with the data within $2\sigma$; and the latter, with $k=2$, is clearly preferred based on our broadband modeling.

Fitting the broadband light curves with $k=2$ and the jet-break time as a free parameter resulted in badly constrained $t_{\rm{j}}$ values above the observed time range ($>238$\,d), and a spherical model seemed to be sufficient. 
When we modeled the data with a fixed $t_{\rm{j}}$ inside the observed time range, we found $\chi^2$ value similar or a bit smaller than those for the spherical model, but this difference was not significant. 
We explored the $t_{\rm{j}}$ parameter space, and found that the $\chi^2$ values were similar for a large range, but for $t_{\rm{j}}<31$\,d the $\chi^2$ value for the jet-break model became larger than the one for the spherical model, and increasingly so for lower $t_{\rm{j}}$ values. 
Therefore, we take $31$\,d as a lower limit on the jet-break time. 
The results of our broadband fits are given in Table~\ref{tab:modres}, and the spherical model fit is shown in Figure~\ref{fig:lcs}. 
We adopted our photometric redshift $z\simeq2.1$ to derive the physical parameters in the table, which we give for both the spherical model and a jet model with $t_{\rm{j}}=31$\,d. 
The values for all the parameters are similar for these two models, and the same is true for $t_{\rm{j}}$ values above 31\,d. 
Since 31\,d is a lower limit on $t_{\rm{j}}$, the jet opening angle $\theta_{\rm{j}}$ and collimation-corrected blast wave energy $E_{\rm{K,jet}}$ in Table~\ref{tab:modres} are also lower limits. 

The parameter values we have found are different than those obtained in broadband modeling work by \citet{zauderer2013}, which we show for comparison in Table~\ref{tab:modres}. 
For some parameters these are only factors of a few, which is not unexpected given the differences in modeling methods. 
Our value for the isotropic equivalent energy $E_{\rm{K,iso}}$, however, is more than one order of magnitude larger than the value they have found, but this could be due to the fact that we are using a larger data set including our $Chandra$, WSRT and AMI data. 
We have a lower limit on the jet-break time which is higher than their value of $t_{\rm{j}}=12$\,d. 
Our lower limit on the jet opening angle, however, is similar to their value due to our higher $E_{\rm{K,iso}}$ value. 
The resulting lower limit we obtained for $E_{\rm{K,jet}}$ is larger than their value by a factor of $\sim3$. 
We note that our value for the density parameter $A_*$, which is $A$ normalized by $5\times10^{11}$, corresponds to a mass loss rate of $\sim4\times10^{-4}$\,M$_{\odot}$/yr for a constant wind velocity of $10^3$\,km/s. 
This mass loss rate is quite typical for Wolf-Rayet stars, usually assumed to be the progenitors of GRBs. 
The mass loss rate, and also the other macro- and microphysical parameters we have derived for \grbnos, are quite typical of what has been found for other GRBs, both for dark \citep[e.g.,][]{jakobsson2005,rol2007,zauderer2013} and optically bright ones \citep[e.g.][]{panaitescu2002,cenko2011b}.

The radio light curves in Figure~\ref{fig:lcs} display a significant scatter around the best fit model, especially at low radio frequencies. 
This can be attributed to the effects of interstellar scintillation (ISS) by free electrons in our galaxy modulating the \grb radio flux \citep[][]{rickett1990}. 
This effect is strongest when the jet angular size is smaller than the characteristic ISS scales, and the modulations quench while the source size increases \citep{goodman1997}. 
To estimate the effects of ISS in the case of \grbnos, we adopt the ``NE2001'' model for the free electrons in our galaxy \citep{cordes2002}, and the scalings for the scintillation strengths and timescales from \citet{walker1998}. 
The coordinates of \grb result in a transition frequency $\nu_0=11$\,GHz between weak and strong scattering, a scattering measure $SM=2.7\times10^{-4}$\,kpc/m$^{20/3}$, and an angular size of the first Fresnel zone $\theta_{\rm{F}0}=\,2.1\,\mu\rm{as}$. 
The resulting predicted scatter due to ISS is indicated by dashed lines in Figure~\ref{fig:lcs}, and it can be seen that the observed scatter can indeed be explained by ISS. 
In fact, for our best fit the $\chi^2_{\rm{red}}$ reduces from a value of 6.2 to 1.9 when ISS is taken into account.

\begin{table}
\begin{center}
\caption{Results from our broadband modeling of \grbnos, for a stellar wind medium ($k=2$, and $A_*$ is $A$ normalized by $5\times10^{11}$). 
The physical parameters are given for a spherical blast wave model and for a model with a jet-break time of 31\,d. 
The latter value for $t_{\rm{j}}$ is the lower limit we determined, and therefore the values for $\theta_{\rm{j}}$ and $E_{\rm{K,jet}}$ are also lower limits. 
We give the parameters obtained in broadband modeling work by \citet{zauderer2013} for comparison.}
\label{tab:modres}
\renewcommand{\arraystretch}{1.1}
\begin{tabular}{|l|c|c|c|}
\hline
Parameter & Spherical & Jet & Zauderer \\
 & model & model & et al. (2013) \\
\hline\hline
$p$ & $2.56$ & $2.57$ & $2.30$ \\
$\varepsilon_{\rm{e}}$ & $7.5\times10^{-2}$ & $8.2\times10^{-2}$ & $1.6\times10^{-1}$ \\
$\varepsilon_{\rm{B}}$ & $3.1\times10^{-5}$ & $4.7\times10^{-5}$ & $9.0\times10^{-5}$ \\
$A_*$ (g/cm) & $38$ & $35$ & $17$ \\
$E_{\rm{K,iso}}$ (erg) & $1.6\times10^{54}$ & $1.4\times10^{54}$ & $7.7\times10^{52}$ \\
$\theta_{\rm{j}}$ (rad) & ... & $>0.14$ & 0.35 \\
$E_{\rm{K,jet}}$ (erg) & ... & $>1.3\times10^{52}$ & $4.6\times10^{51}$ \\
\hline
\end{tabular}
\end{center} 
\end{table}

\section{Optical Darkness}\label{sec:darkness}

\begin{table*}
\scriptsize
\begin{center}
\caption{Observed limits on the magnitude and flux density compared to model flux calculations: lower limits on the extinction, and on the $E_{B-V}^{\rm{host}}$ and $A_{V}^{\rm{host}}$ values, based on the model flux calculations (after correcting for Galactic extinction) and for Galactic, LMC and SMC extinction curves; and upper limits on the optical-to-X-ray spectral index $\beta_{\rm{OX}}$. The $K_s$, $J$ and $z'$ band measurements are from this paper, the $i$ band measurement is taken from \citet{zauderer2013}.}
\label{tab:extinct}
\renewcommand{\arraystretch}{1.1}
\begin{tabular}{|l|c|c|c|c|c|c|c|c|c|c|c|} 
\hline
Filter & $\Delta T$ & Magni- & Flux & $\beta_{\rm{OX}}$ & Extinc- & Galactic & Galactic & LMC & LMC & SMC & SMC \\
 & (days) & tude & density & & tion & $E_{B-V}^{\rm{host}}$ & $A_{V}^{\rm{host}}$ & $E_{B-V}^{\rm{host}}$ & $A_{V}^{\rm{host}}$ & $E_{B-V}^{\rm{host}}$ & $A_{V}^{\rm{host}}$ \\
 & & & ($\mu$Jy) & & (mag) & & (mag) &  & (mag) &  & (mag) \\
\hline\hline
$K_s$ & 0.21 & $>19.2$ & $<13$ & $<0.34$ & $>6.1$ & $>2.3$ & $>7.5$ & $>1.8$ & $>6.1$ & $>2.8$ & $>7.6$ \\
$J$ & 0.22 & $>19.5$ & $<7.7$ & $<0.22$ & $>4.9$ & $>1.1$ & $>3.7$ & $>1.1$ & $>3.8$ & $>1.3$ & $>3.5$ \\
$i$ & 0.16 & $>21.7$ & $<8.4$ & $<0.20$ & $>6.7$ & $>0.9$ & $>3.0$ & $>1.0$ & $>3.4$ & $>1.1$ & $>3.0$ \\
$z'$ & 0.29 & $>22.6$ & $<3.9$ & $<0.22$ & $>7.6$ & $>1.4$ & $>4.5$ & $>1.2$ & $>4.2$ & $>1.4$ & $>3.8$ \\
\hline
\end{tabular}
\end{center}
\end{table*}

\grb was not detected down to deep limits at nIR, optical, and UV frequencies. 
In the literature there are three methods to establish if a GRB can be classified as a dark burst. 
The first method by \citet{jakobsson2004} uses the simultaneous optical and X-ray flux, at roughly half a day after the GRB onset, and requires an optical-to-X-ray spectral index $\beta_{\rm{OX}}<0.5$ for a GRB to be classified as dark. 
A more elaborate approach, using information from both spectra and light curves, was developed by \citet{rol2005}. 
These two methods make strong assumptions about the underlying physical model, namely that the electron energy distribution index $p$ has to be larger than 2 \citep{jakobsson2004}, or that the dynamics of GRB jets is well understood and can be described by simplified analytic models \citep{rol2005}. 
A third method, proposed by \citet{vanderhorst2009}, only uses spectral information, makes fewer assumptions on the GRB physics, and bases the classification on a comparison of the simultaneous X-ray and optical-to-X-ray spectral index. 
In the latter method a GRB is classified as dark when $\beta_{\rm{OX}}<\beta_{\rm{X}}-0.5$, with $\beta_{\rm{X}}$ the X-ray spectral index. 

As already mentioned in Section~\ref{sec:intro}, the main assumption of all classification methods is that the optical and X-ray emission are part of the same spectrum, and thus that there is only one emission process at play, namely synchrotron emission from the GRB blast wave moving into the ambient medium. 
Since prompt emission, originating at a different emission site to the forward shock, was detected at gamma-ray and X-ray frequencies up to at least half an hour after the GRB onset, none of the early-time optical upper limits should be considered for the optical classification. 
Therefore, we only use the observations after a few hours for this purpose. 
In Table~\ref{tab:extinct} we give the most sensitive observations of \grb on this timescale \citep[Section~\ref{sec:obs};][]{zauderer2013}. 
Based on the upper limits on the nIR/optical flux and the (quasi-)simultaneous X-ray fluxes, we determined the optical-to-X-ray spectral indices, which are fairly similar for these four observations, and the most constraining value is $\beta_{\rm{OX}}<0.20$. 
Since the X-ray spectral index is $\beta_{\rm{X}}\simeq1.0$ (Section~\ref{sec:obs}), $\beta_{\rm{OX}}-\beta_{\rm{X}}<-0.8$. 
Therefore, \grb is clearly a dark burst, according to the classification methods of both \citet{jakobsson2004} and \citet{vanderhorst2009}. 

Since we have determined the redshift of \grb to be $z\simeq2.1$, a very high redshift is not the cause of its darkness. 
We have also established that the light curves at a few hours can be modeled well as synchrotron emission from the forward shock, and thus the remaining viable explanation for the optical darkness is extinction in the host galaxy. 
The results of our modeling work (Section~\ref{sec:modeling}) can be used to estimate the expected flux density at the times of the observations listed in Table~\ref{tab:extinct}. 
Comparing these model flux densities with the observed upper limits, we estimate lower limits on the extinction (after correcting for Galactic extinction), which range from 4.9 to 7.6 mag (Table~\ref{tab:extinct}). 
In this table we also give estimates for $E_{B-V}^{\rm{host}}$ and $A_{V}^{\rm{host}}$ for three types of extinction curves: Galactic \citep{cardelli1989}, SMC and LMC \citep{gordon2003}. 
This assumes that one of these extinction curves is a valid model for the host galaxy of \grbnos, or more specifically for the region causing the optical extinction. 
Therefore, these $E_{B-V}^{\rm{host}}$ and $A_{V}^{\rm{host}}$ values should be taken with caution, but can be used for comparison with other GRB studies. 
For a Galactic extinction curve the highest lower limit is $E_{B-V}^{\rm{host}}>2.3$, corresponding to $A_{V}^{\rm{host}}>7.5$, which is similar to the value found by \citet{zauderer2013}, and much higher than the average value for the host galaxy ($E_{B-V}\simeq0.15$; Section~\ref{sec:photoz}). 
The latter indicates that there is a dusty region in the host galaxy in our line of sight towards \grbnos, which is consistent with the lack of bright emission at the GRB location in our {\it HST} images (Section~\ref{sec:hst}), and with the findings for other dark bursts for which similar modeling work has been performed \citep{jakobsson2005,rol2007,zauderer2013}. 
Note that the best extinction limits are for our $K_s$ observation, which has the weakest constraint on $\beta_{\rm{OX}}$ ($<0.34$). 
This illustrates how important deep nIR observations are, with 4~m class telescopes like WHT, to constrain the extinction in the galactic environment of GRBs.

Finally, we estimate the expected $A_{V}^{\rm{host}}$ based on the excess absorption $N_{\rm{H,host}}$ at the \grb redshift compared to the Galactic absorption $N_{\rm{H,gal}}$ in the X-ray spectra. 
We adopted two approaches, resulting in a lower limit and an illustrative estimate for $A_{V}^{\rm{host}}$. 
To obtain the lower limit we have fit the combined {\it Swift}/XRT Photon Counting mode spectrum over $0.3-10$\,keV assuming the 90\% lower limit on the redshift of $z=1.85$, $N_{\rm{H,gal}}$ from \citet{willingale2013}, and a Solar metallicity absorber in the host galaxy. 
This resulted in $N_{\rm{H,host}}>3.4\times10^{22}$\,cm$^{-2}$ (90\% confidence level), which corresponds to $A_{V}^{\rm{host}}>15$ and $E_{B-V}^{\rm{host}}>4.7$ (90\% as well) using the results from \citet{guver2009}. 
We also determined $N_{\rm{H,host}}$ from our photometric redshift $z=2.06$ adopting LMC metallicity \citep{pei1992}, which is a more realistic metallicity for GRB hosts \citep[e.g.,][]{schady2012}. 
We find $N_{\rm{H,host}}=(1.1\pm0.2)\times10^{23}$\,cm$^{-2}$, and thus $A_{V}^{\rm{host}}=50\pm9$ and $E_{B-V}^{\rm{host}}=15\pm3$. 
It is clear that these limits and estimates on $A_{V}^{\rm{host}}$and $E_{B-V}^{\rm{host}}$ are consistent with the findings from our broadband modeling. 
Our estimates are higher than those of \citet{zauderer2013}, but within a factor of 2, which can be explained by adopting slightly different methods to obtain these values.

\section{Conclusions}\label{sec:conclusions}

\grb was bright at radio and X-ray frequencies, but not detected in the nIR/optical/UV bands, and we have shown that it clearly belongs to the class of dark bursts. 
This is one of the few dark bursts with well-sampled X-ray and radio light curves at multiple observing frequencies, which allowed detailed broadband modeling. 
Our {\it Chandra} X-ray observation complements the {\it Swift}/XRT light curve to better constrain the lower limit on the jet-break time. 
Together with our WSRT and AMI light curves at three radio frequencies, and radio observations presented in the literature, we have found that the \grb data can in fact be modeled well with a spherical blast wave moving into a circumburst medium structured like a stellar wind. 
From this broadband modeling we put a lower limit on the jet-break time of 31\,d, and find that all the other macro- and microphysical parameters of \grb are typical of what has been found for other, optically dark and bright, GRBs. 

We have combined our broadband modeling results with deep nIR observations from a few hours after the GRB onset to constrain the extinction in the host galaxy of \grbnos, the most likely cause of the optical darkness of this GRB. 
Our $K_s$ band observation at $\sim5$\,h puts the most stringent constraints, namely $E_{B-V}^{\rm{host}}>2.3$ and $A_{V}^{\rm{host}}>7.5$ if one adopts a Galactic extinction curve for the host galaxy. 
These limits are consistent with the optical extinction we estimated based on the excess absorption in the X-ray spectra. 
From Keck and {\it Spitzer} imaging of the host we have determined the photometric redshift $z_{\rm{phot}}\simeq2.1$, and the stellar mass and star formation rate of that galaxy, which are both typical for dark GRBs. 
Finally, by combining deep {\it HST} images of the host galaxy with our {\it Chandra} image we have shown that the GRB position lies off-set from any bright regions of star formation in the host galaxy. 
We suggest that either the \grb progenitor was formed in a region of relatively low star formation, or that this region is so heavily obscured by gas or dust that it is even not visible in the nIR.

\section*{Acknowledgments}
We would like to thank Alexander Kann for useful discussions, and Alex Filippenko, Shri Kulkarni, Josh Bloom, Brad Cenko, and Jeff Silverman for enabling or performing the Keck telescope observations presented in this paper. 
AJvdH and RAMJW acknowledge support from the European Research Council via Advanced Investigator Grant no. 247295. 
KW acknowledges support from the Science \& Technology Facilities Council (STFC). 
Support for DAP was provided by NASA through Hubble Fellowship grant HST-HF-51296.01-A awarded by the Space Telescope Science Institute, which is operated by the Association of Universities for Research in Astronomy, Inc., for NASA, under contract NAS 5-26555. 
RLCS is supported by a Royal Society Fellowship. 
PAC acknowledges support from Australian Research Council grant DP120102393. 

The Westerbork Synthesis Radio Telescope is operated by ASTRON (Netherlands Institute for Radio Astronomy) with support from the Netherlands foundation for Scientific Research. 
The Arcminute Microkelvin Imager arrays are supported by the University of Cambridge and the STFC. 
The William Herschel Telescope and Nordic Optical Telescope are operated on the island of La Palma by the Isaac Newton Group and Nordic Optical Telescope Scientific Association, respectively, in the Spanish Observatorio del Roque de los Muchachos of the Instituto de Astrofisica de Canarias. 
Some of the data presented herein were obtained at the W.M. Keck Observatory, which is operated as a scientific partnership among the California Institute of Technology, the University of California and the National Aeronautics and Space Administration (NASA); the Observatory was made possible by the generous financial support of the W.M. Keck Foundation. 
This work is based in part on observations made with the {\it Spitzer Space Telescope}, which is operated by the Jet Propulsion Laboratory, California Institute of Technology, under a contract with NASA; {\it Spitzer} observations were undertaken as part of large program 90062. 
Some observations were made with the NASA/ESA {\it Hubble Space Telescope}, obtained at the Space Telescope Science Institute, which is operated by the Association of Universities for Research in Astronomy, Inc., under NASA contract NAS 5-26555; {\it HST} observations were undertaken as part of program 12378. 
The scientific results reported in this article are based in part on observations made by the {\it Chandra} X-ray Observatory, under ObsID 14052. 
This work made use of data supplied by the UK Swift Science Data Centre at the University of Leicester.

\bibliographystyle{mn2e}
\bibliography{references}

\end{document}